\documentclass{article}
\usepackage[accepted]{vietnam} 
\usepackage{natbib}
\usepackage{graphicx}      
\usepackage{color}
\usepackage{hyperref} 
\hypersetup{colorlinks=true,citecolor=blue}

\begin{document}
\twocolumn[
\title{From Clumps to Cores to Protostars: The JCMT Gould Belt Survey's ``First-Look'' analysis of Southern Orion A using SCUBA-2}
\titlerunning{GBS First Look at Southern Orion A}
\author{S. Mairs$^{1,2}$, D. Johnstone$^{2}$, H. Kirk$^{2}$}{smairs@uvic.ca}
\address{$^{1}$University of Victoria, 3800 Finnerty Road, Victoria, BC, Canada, V8P 5C2}
\address{$^{2}$Herzberg Astrophysics, National Research Council of Canada, 5071 West Saanich Road, Victoria, BC, V8N 3L3}

\vskip 0.5cm 
]

\begin{abstract}
We present a subset of the James Clerk Maxwell Telescope Gould Belt Survey's ``first-look'' results of the southern extent of the Orion A molecular cloud. Employing a two-step structure identification process, we constructed individual catalogues for large-scale regions of significant emission labeled as ``islands'' and smaller-scale subregions called ``fragments'' using the 850 $\mu$m continuum maps obtained using the Submillimetre Common-User Bolometer Array 2 (SCUBA-2). We highlight the relationship between the concentration and Jeans stability for regions of significant emission and present the results of an investigation into the spatial distribution of Young Stellar Objects  detected using the \textit{Spitzer Space Telescope} and the \textit{Herschel Space Observatory}. We find an apparent evolution in the velocity dispersion from Class 0 to Class II objects which we derive from comparing our observations to a simple model.
\end{abstract}

\section{Introduction}

The James Clerk Maxwell Telescope's (JCMT) Gould Belt Legacy Survey (GBS, \citealt{Wt2007}) is a large-scale project which has mapped the notable star-forming regions within 500 pc of the Sun.
In these proceedings, we present a subset of the first results from the Southern Orion A region (\citealt{Mairs2016}) observed at \mbox{\mbox{850 $\mu$m}} with the Submillimetre Common-User Bolometer Array 2 (SCUBA-2) instrument (\citealt{Holland2013}). 

The detected emission in Southern Orion A includes several active sites of Galactic star formation such as OMC-4, OMC-5, and L1641N. It contains dozens of embedded sources (\citealt{Johnstone2006}; \citealt{Ali2004}; \citealt{Chen1996}), 
as well as several Herbig-Haro objects. 
The SCUBA-2 observations presented in \citet{Mairs2016} have a sensitivity which is an order of magnitude deeper than previous maps (see \citealt{Johnstone2006}) along with a much wider spatial coverage (8100 arcmin$^{2}$ compared to 2300 arcmin$^{2}$ in the original Southern Orion A SCUBA data).  Thus, we have a much better diagnostic to characterise the dense, cold dust.  To complement these new continuum observations of dense, often gravitationally unstable gas, we use extinction data taken in the \textit{J}, \textit{H} and \textit{K} bands (using the Two Micron All Sky Survey, 2MASS) that were determined by the Near-infrared Color Excess ({\sc{NICE}}) team (M. Lombardi, private communication, July 18$^{th}$, 2015), and the young stellar object (YSO) catalogues of \citet{Megeath2012} and \citet{Stutz2013} obtained using the \textit{Spitzer Space Telescope} and the \textit{Herschel Space Observatory}, respectively.

In Section \ref{structuresec}, we present our structure identification procedure and highlight some of the key results with regard to the continuum emission. In Section \ref{ysoresultssec}, we discuss the spatial distribution of Young Stellar Objects (YSOs) and compare this result to a simple model assuming the YSOs were launched from dense cores currently showing evidence of star formation. 

\section{Fragmentation of Dense Structures}
\label{structuresec}

Southern Orion A contains a diverse set of objects defined by localized emission. 
We consider a pixel to be ``significant'' if it has a value of at least 3$\sigma_{rms}$ (\mbox{$\sigma_{rms} = 3.1\mathrm{\:mJy\:beam}^{-1}\mathrm{\:}$}) in the \mbox{850 $\mu$m} map. We first extract the largest objects studied in this work by simply drawing a contour at 3$\sigma_{rms}$  and retaining all enclosed structures larger than approximately one beam  (15'' in circularly projected diameter). We accomplish this identification using Starlink's version of the algorithm {\sc{ClumpFind}} (\citealt{Williams1994}) as implemented in the {\sc{Cupid}} package (\citealt{Berry2007}) by defining only one flux level over which significant structure is identified. Each non-spurious object detected is referred to as an ``\textit{island}''; any flux present in the map outside of an island is considered to be dominated by noise. 

In the second step, we employ the JCMT Science Archive algorithm {\sc{jsa\_catalogue}} found in Starlink's {\sc{PICARD}} package (\citealt{Gibb2013}). This algorithm uses the {\sc{FellWalker}} routine (\citealt{Berry2015}). Briefly, {\sc{FellWalker}} marches through a given image pixel by pixel and identifies the steepest gradient up to an emission peak. After performing tests to ensure that the peak is ``real'' and not just a noise spike, the local maximum is assigned an identifying integer and all the pixels above a user-defined threshold that were included in the path to the peak are given the same identifier. In this way, all of the robust peaks in the image are catalogued and the structure associated with each peak can be analyzed. These localized peaks often separate emission contained within the larger islands into multiple components. In this way, the compact source catalogue generated reveals the substructure present within the context of coincident large-scale emission. For this reason, we label the compact components as ``\emph{fragments}''. We refer to an island which contains at least two fragments as a ``complex island'' and an island that contains only one fragment as a ``monolithic island''.

\subsection{Stability and Concentration}

\begin{figure}
\vskip -0.3cm
\centering
$\begin{array}{cc}
\includegraphics[width=8cm,height=7cm]{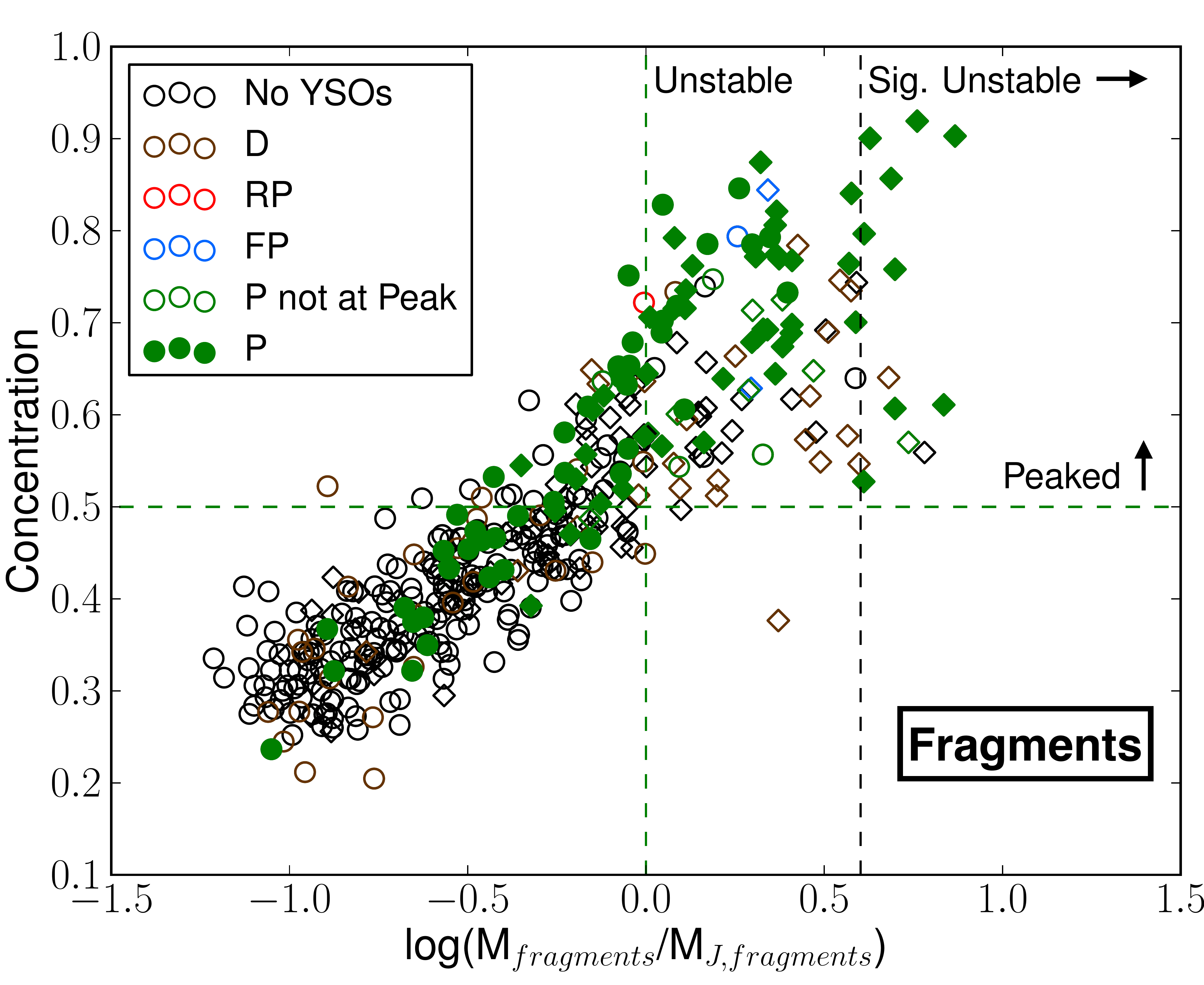} 
\end{array}$
\caption{\it Fragment concentration versus fragment stability. The dashed green lines show a concentration of 0.5 on the ordinate and the gravitational instability line on the abscissa. The vertical dashed black line represents an $M/M_{J}$ ratio of 4 where we define sources to be significantly unstable. Colours represent associations between the identified fragment and several classes of YSOs as denoted in the legend. Diamonds represent a fragment which belongs to a complex island and a circle represents a fragment which traces isolated, monolithic structure. This figure is taken from \citet{Mairs2016}. 
}
\label{concenstab}
\end{figure}

More so than islands, it is the compact, localized fragments for which we expect Jeans unstable cases to be forming (or to eventually go on to form) stars. Thus, in Figure \ref{concenstab}, we compare fragment concentrations with their Jeans stabilities. The concentration, $C$, is a useful metric to quantify whether or not a structure is peaked. The concentration is calculated by comparing the total flux density measured across a given island or fragment to a uniform structure of the same area wherein each pixel is set to the peak brightness, $f_{850,peak}$. Following \citet{Johnstone2001}, $C = 1 - \frac{1.13B^{2}S_{850}}{\pi R^{2}\times f_{850,peak}}$, where $B$ is the beam width in arcseconds, $R$ is the radius of the source measured in arcseconds, $S_{850}$ is the total flux of the source measured in Jy, and $f_{850,peak}$ is the peak brightness of the source measured in \mbox{Jy beam$^{-1}$}. Highly concentrated sources are expected to have a higher degree of self-gravity (see \citealt{Johnstone2001} and \citealt{Kirk2006}), eventually collapsing and forming one to a few stellar systems. To calculate the Jeans stability, we compare the mass of a fragment, $M$, to its associated Jeans mass, $M_{J}$ 
(\citealt{Mairs2016}).

In Figure \ref{concenstab}, each fragment is colour-coded by its association with YSOs identified in the \citet{Megeath2012} and \citet{Stutz2013} catalogues. Class 0+I and flat spectrum sources are denoted ``P'' for protostar. A green outline is given for protostars contained within a fragment's boundaries and a solid green point is given for fragments which have a protostar within 15'' (1 beam) of the peak flux measurement. Class II+III objects are labelled ``D'' for disk sources. Red (``RP'') and blue (``FP'') outlines are given for two classes of protostar candidates, see \citealt{Megeath2012}).  Green dashed lines indicate the nominal gravitational instability line \mbox{$M/M_{J} \geq $ 1} (horizontal) and  $C = 0.5$ (vertical). $C = 0.5$ is chosen because it represents a relatively concentrated core approximately half way between a uniform density (0.33) and self gravitating Bonnor Ebert sphere (0.72) (see \citealt{Johnstone2001}). Note that the fragments fall broadly into two regimes: 1. gravitationally stable and low concentration and 2. gravitationally unstable and with peaked emission. 
We note as well that the diamond symbols in Figure \ref{concenstab} represent a fragment which belongs to a complex island (an island containing at least two fragments) and a circle represents a fragment which traces isolated, monolithic structure. We would expect the gravitationally unstable, peaked fragments to be the population which is associated with protostars. In general, we see that this is the case. In Figure \ref{concenstab}, only 8\% of the fragments without discernible signs of YSOs appear unstable and concentrated. Of those, the fragments which were extracted from monolithic islands (or have no island associations) are outnumbered by those which were extracted from complex islands (21\% and 79\%, respectively). Conversely, we would expect the gravitationally stable, less peaked fragments to be the population which is not actively forming stars. Indeed, only 23\% of the  stable and uniform fragments appear to have YSOs. Almost all of these fragments are associated with monolithic islands (83\%); that is, they do not have ``siblings'' within the same island. A catalogue of interesting follow-up source candidates is presented in \citet{Mairs2016}.

 \begin{figure*}[!htbp]
\vskip -1cm
\centering
$\begin{array}{c}
\hspace{-0.5cm}
\includegraphics[width=17.5cm]{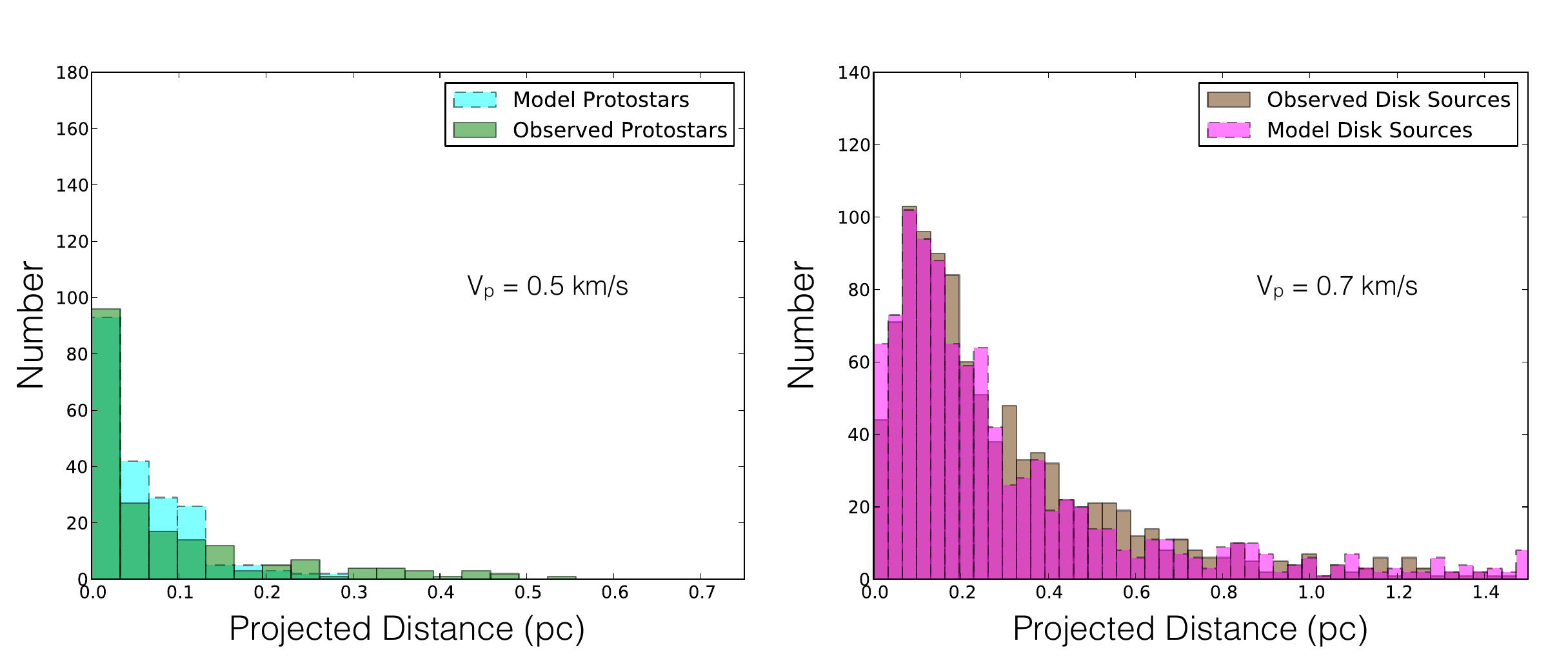} 
\end{array}$
\caption{\it {\bf Left:} The calculated projected distance between model protostar locations and the nearest fragment peak brightness location assuming $v_{p} = 0.5 \mathrm{\:km\:s}^{-1}$  (cyan, dashed lines) plotted along with the observed distribution (green, solid lines). {\bf Right:} The calculated projected distance between model disk source locations and the nearest fragment peak brightness location assuming $v_{p} = 0.7 \mathrm{\:km\:s}^{-1}$ (magenta, dashed lines) plotted along with the observed distribution (brown, solid lines). This figure is taken from \citet{Mairs2016}. 
}
\label{ysodistfig}
\end{figure*}

\section{The Young Stellar Object Distribution}
\label{ysoresultssec}

An analysis of the spatial distribution of YSOs with respect to the location of the nearest fragment's peak emission is presented in \citet{Mairs2016}. In that analysis, it is clear that the surface densities of these sources can be separated into two populations we label as ``clustered'' (away from the edges of the map and close to fragments) and ``distributed'' (the sporadic sources at larger distances from the clustered objects around fragments). We attempted to recreate both populations using a simple model assuming: 1. 
 the currently observed structures are linked to the formation of young stars and their present distribution, 2. all of the observed YSOs formed in fragments which are calculated to be Jeans unstable and every Jeans unstable fragment has the same probability of producing a YSO, 3. the half-life age of disks is estimated to be $t_{0.5}= 2 \mathrm{\:Myr}$ and we detect no discs older than 10 Myr, 4. protostars to have an age $\leq 0.5$ Myr, and 5. the random 3D space velocities of the observed YSOs follow a Maxwell-Boltzmann distribution with a fixed most probable speed, $v_{p}$. Ten $v_{p}$ values were tested from 0.1 km s$^{-1}$ to 1.0 km s$^{-1}$.

In order to recreate both the clustered and distributed populations of YSOs simultaneously, it was necessary to fit different $v_{p}$ values to the different YSO classes (younger protostars and more evolved disk sources). The left panel of Figure \ref{ysodistfig} shows a comparison between the observed projected distance between protostar (Class 0+I and flat spectrum sources) locations and the nearest fragment peak brightness location  and the results obtained from our model assuming $v_{p} = 0.5 \mathrm{\:km\:s}^{-1}$. We found that the $v_{p}$  value which best fits the protostar population between 0.2 km s$^{-1}$ and \mbox{0.5 km s$^{-1}$}.  The right panel of Figure \ref{ysodistfig} shows the same results for the disk sources (Class II+III) assuming the best fitting $v_{p}$ value of  $0.7 \mathrm{\:km\:s}^{-1}$. Note that \citet{Jorgensen2007}, through observations of the Perseus molecular cloud, and \citet{Frimann2016}, through the MHD simulation {\sc RAMSES}, found the velocity dispersion of Class 0 objects to be $\sim$0.1-0.2 km s$^{-1}$. Thus, there appears to be a trend in the velocity with YSO class. These velocities, however, are highly dependent on the lifetimes of each type of object.

\vspace{0.3cm}

For more information on this work, including further analyses and full catalogues of the observed structures, see \citet{Mairs2016}.

\section*{Acknowledgments}
\vspace{-0.3cm}
Steve Mairs was partially supported by the Natural Sciences and
Engineering Research Council (NSERC) of Canada graduate scholarship
program. Doug Johnstone is supported by the National Research Council
of Canada and by an NSERC Discovery Grant. The authors wish to thank 
ICISE for hosting SFDE 2016 and extend our gratitude to both the LOC and SOC
for organising this conference.


\end{document}